\documentclass{PoS}
\title{Cherenkov light from Horizontal Air Shower}

\ShortTitle{Horizontal EAS simulations}

\author{\speaker{K. Kr\'{o}lik} $^{a, b}$, A. Djakonow $^{a,c}$,  Z. Plebaniak $^{a}$,  M. Przybylak $^{a}$, J. Szabelski $^{a}$, L.~Wiencke $^{d}$  for the JEM-EUSO Collaboration\\
        \llap{$^a$}National Centre for Nuclear Research, Astrophysics Division, Cosmic Ray Laboratory, \\ul. 28 Pu{\l }ku Strzelc\'ow Kaniowskich 69, 90-558 {\L }\'od\'z, Poland.
        \\
	\llap{$^b$} Faculty of Physics, University of Warsaw, ul. Pasteura 5, 02-093 Warsaw, Poland. \\
	\llap{$^c$} Faculty of Technical Physics, Information Technology and Applied Mathematics, {\L }\'od\'z University of Technology, ul. W\'olcza\'nska 215, 90-924, {\L }\'od\'z, Poland.\\
        \llap{$^d$}  Colorado School of Mines\\
	E-mail: \email{krolik.karolina1@gmail.com}	
        }

\abstract{We present results of horizontal EAS simulations focused on the opportunity of measuring Cherenkov light from air showers at stratospheric balloon altitude (eg. EUSO-SPB2). For a\,$1\,\mathrm{m^2}$ UV light detector at a 38 km altitude, the largest horizontal distance to the edge of the Earth atmosphere is about 1000 km which represents a depth of $10000\,\mathrm{g/cm^2}$ of atmosphere. The Cherenkov light produced by the EAS electron component would be scattered in atmosphere on its way to the detector, and would not contribute to detected light. The most promising scenario relies on the detection of light emitted within about 300 km from the detector by EAS muons with energies above 100 GeV (required to produce Cherenkov light at high altitudes and for muons to survive over a large distance). Within this scenario we might expect to measure Cherenkov light from proton induced EAS of energy between $10^{17}$ and $10^{18}\,\mathrm{eV}$, the lower limit being related to the strength of a signal, and upper limit being due to the product of geometrical factor by the CR flux.}

\FullConference{36th International Cosmic Ray Conference -ICRC2019-\\
		July 24th - August 1st, 2019\\
		Madison, WI, U.S.A.}

\begin{document}

\section{Introduction}

Balloon flight gives unique opportunity to observe development of Extensive Air Showers (EAS), generated by Ultra High Energy Cosmic Rays (UHECR). Previous flight -- EUSO-SPB (Extreme Universe Space Observatory - Super Pressure Balloon) -- started in April 2017 from Wanaka, New Zealand \cite{Wiencke:2017cfi}. Detector was attached under the balloon and looking down to the atmosphere to detect EAS by observing UV light during nights. Next JEM-EUSO mission's detector -- EUSO-SPB2 \cite{Adams:2017fjh} -- will be looking at the horizon what gives new observational opportunities. One is to observe, for the first time, EAS developing nearly horizontally in the atmosphere, both by the UV and direct Cherenkov light. 

This publications describe results of estimations the possibility of detecting horizontal EAS, developing directly into detector, by UV Cherenkov light in wavelength range 290 nm - 400 nm emitted mostly by high energy muons.

\section{Simulations problems}
\begin{figure}
\centering
\includegraphics[scale=0.5]{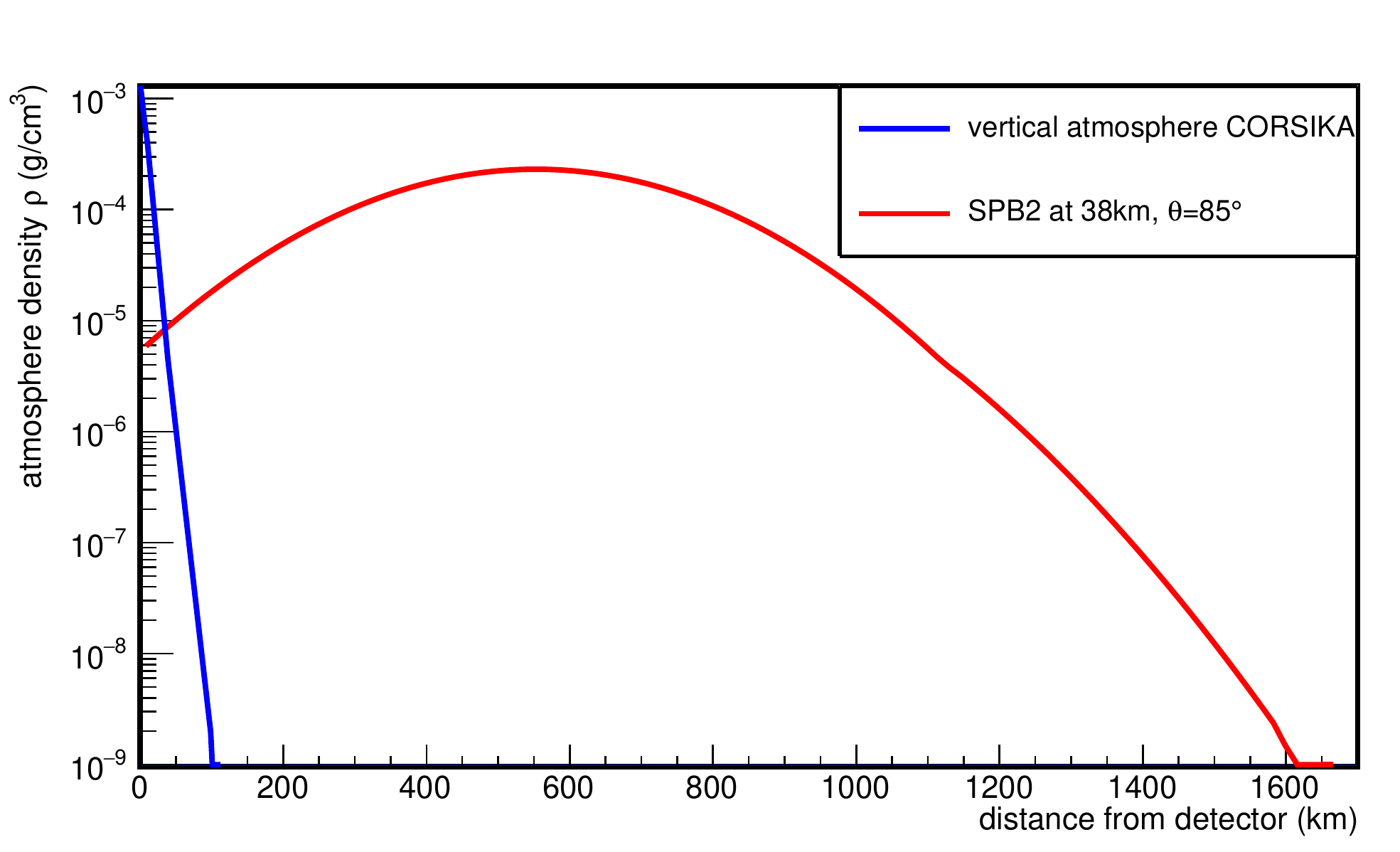}
\caption{Atmosphere profiles based on U. S. Standard Atmosphere from CORSIKA, blue: vertical profile, red: profile for horizontal EAS developing at detector angle $\psi = 85^{\circ}$ at 38 km altitude.}
\label{fig:atmosphere_density_profile}

\end{figure}

One possible way to obtain the number of Cherenkov photons reaching EUSO detector is to run Monte Carlo simulations using CORSIKA code. In CORSIKA atmosphere is described with density profile shown in fig. \ref{fig:atmosphere_density_profile} called U. S. Standard Atmosphere. It can be also described by thickness $T$ connected with density $\rho$ as follows:
\begin{equation}
T_x = \int_{h_{top}} ^{h_{x}} \rho\left(h(l)\right) \mathrm{d}l,
\end{equation}
where $h_{top}$ is at the edge of atmosphere (112.8 km) and $h_x$ is a height of position; $T_x$ is thickness at that position as integral along particle trajectory in the atmosphere

\begin{figure}
\centering
\includegraphics[scale=0.5]{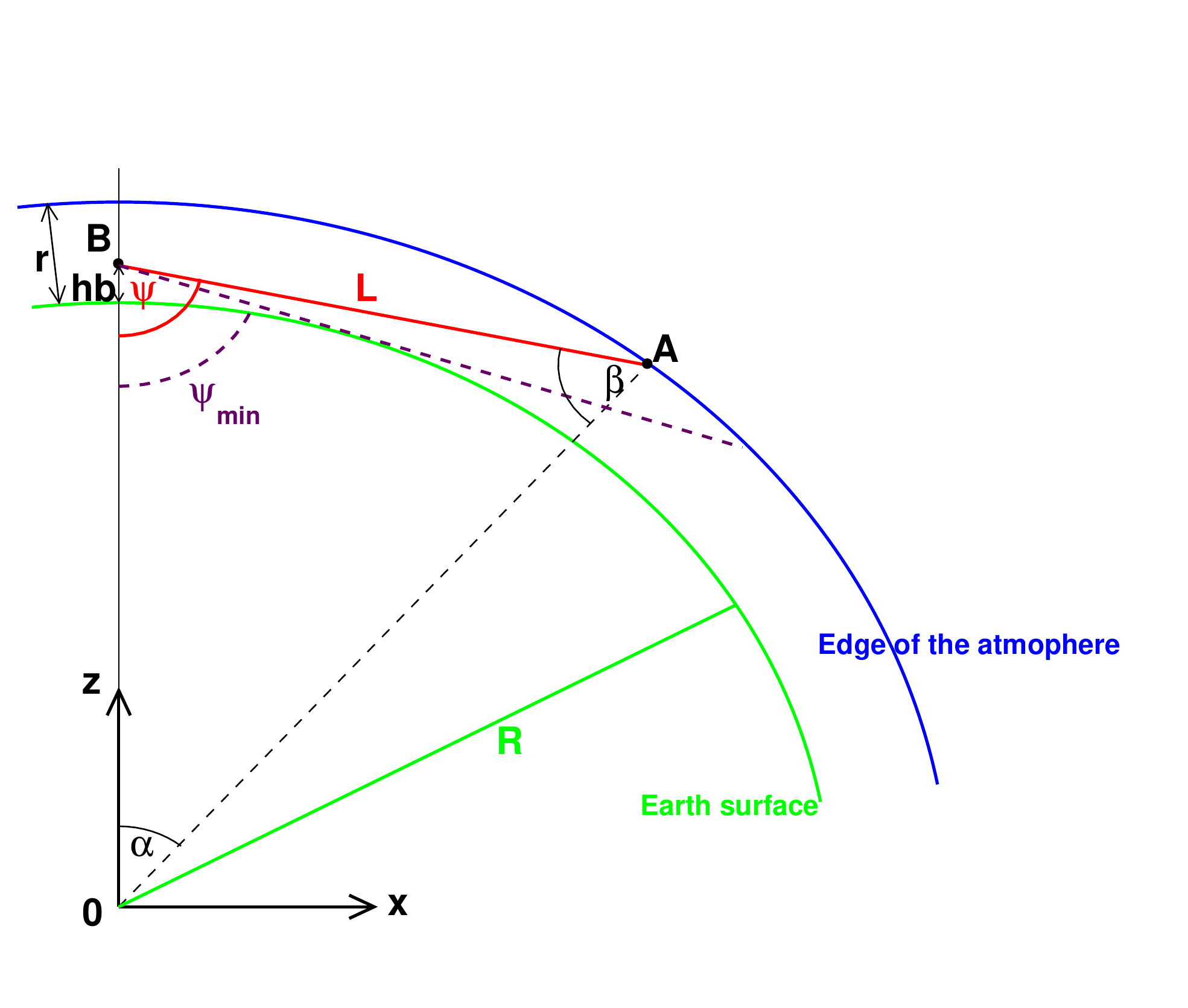}
\caption{Coordinate system used in our considerations. }
\label{fig:coordinate_system}

\end{figure}

\begin{figure}
\centering
\includegraphics[scale=0.5]{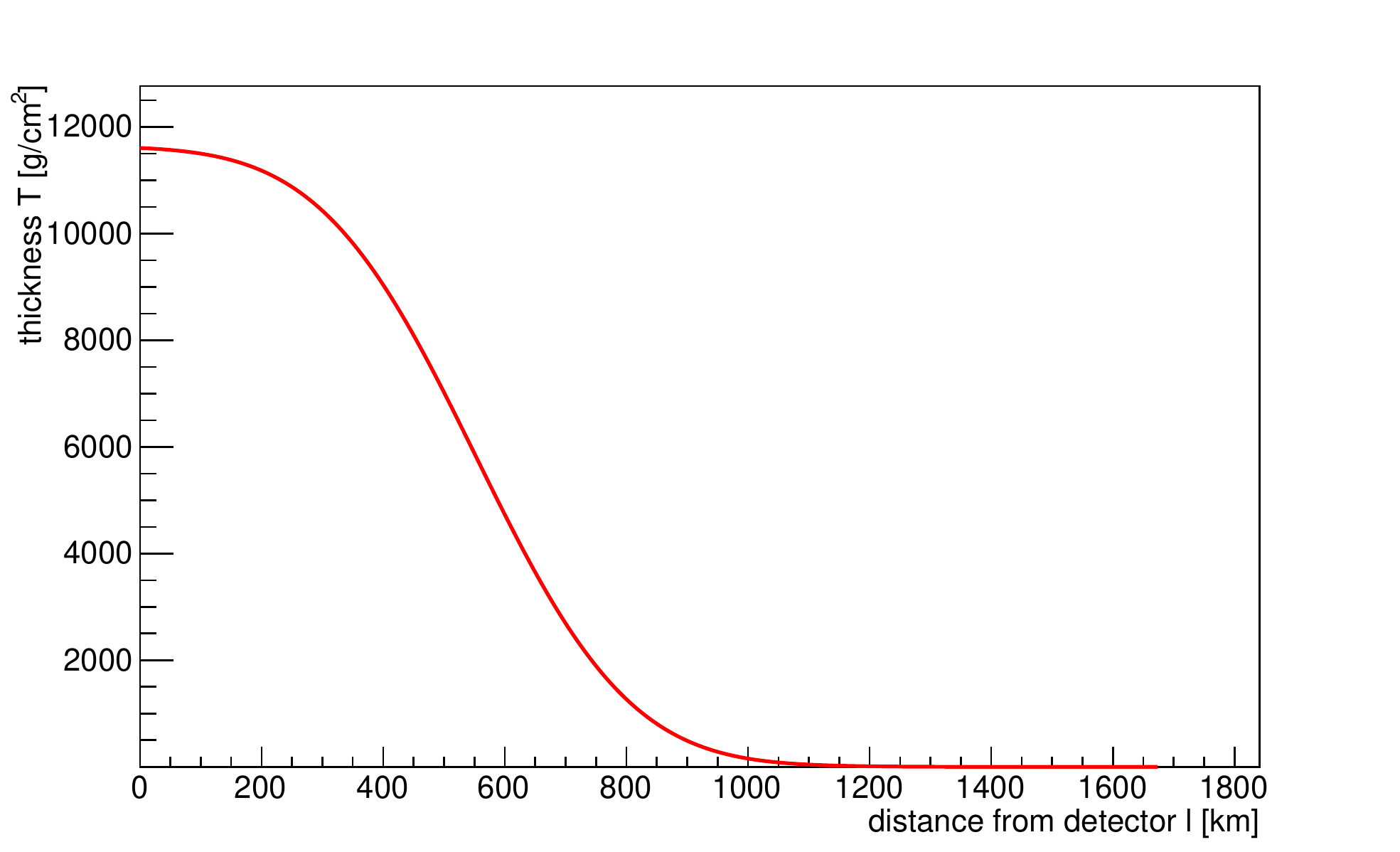}
\caption{Atmosphere thickness profile for horizontal EAS developing at detector angle $\psi = 85^{\circ}$ at 38 km altitude.}
\label{fig:atmosphere_thickness_profile}

\end{figure}

In our case UHECR enters the atmosphere in point $A$ and develops EAS along line $L$ to the detector in point $B$ (fig. \ref{fig:coordinate_system}). Atmosphere profile described by density $\rho$ on the way of EAS development is shown in fig. \ref{fig:atmosphere_density_profile}. For the SPB2 detector at $38\,\mathrm{km}$ altitude tilted $85^{\circ}$ from normal, distance between points $A$ and $B$ are about $1600\,\mathrm{km}$ and $12000\,\mathrm{g/cm^2}$ of the atmosphere (fig. \ref{fig:atmosphere_thickness_profile}). It is ten times more than for vertically developing EAS, so shower will fully develop hundreds of kilometers before the detector. Even for the highest energies, proton induced vertical EAS have maximum before thickness $1000\,\mathrm{g/cm^2}$  of atmosphere. From atmosphere profile from fig. \ref{fig:atmosphere_thickness_profile} one can calculate that $1000\,\mathrm{g/cm^2}$ are about 800 km from the edge of the atmosphere for AB trajectory. In the fig. \ref{fig:atmosphere_density_profile} we can see that after this distance along particle trajectory, atmosphere is much more dense. 

\begin{figure}
\centering
\includegraphics[scale=0.5]{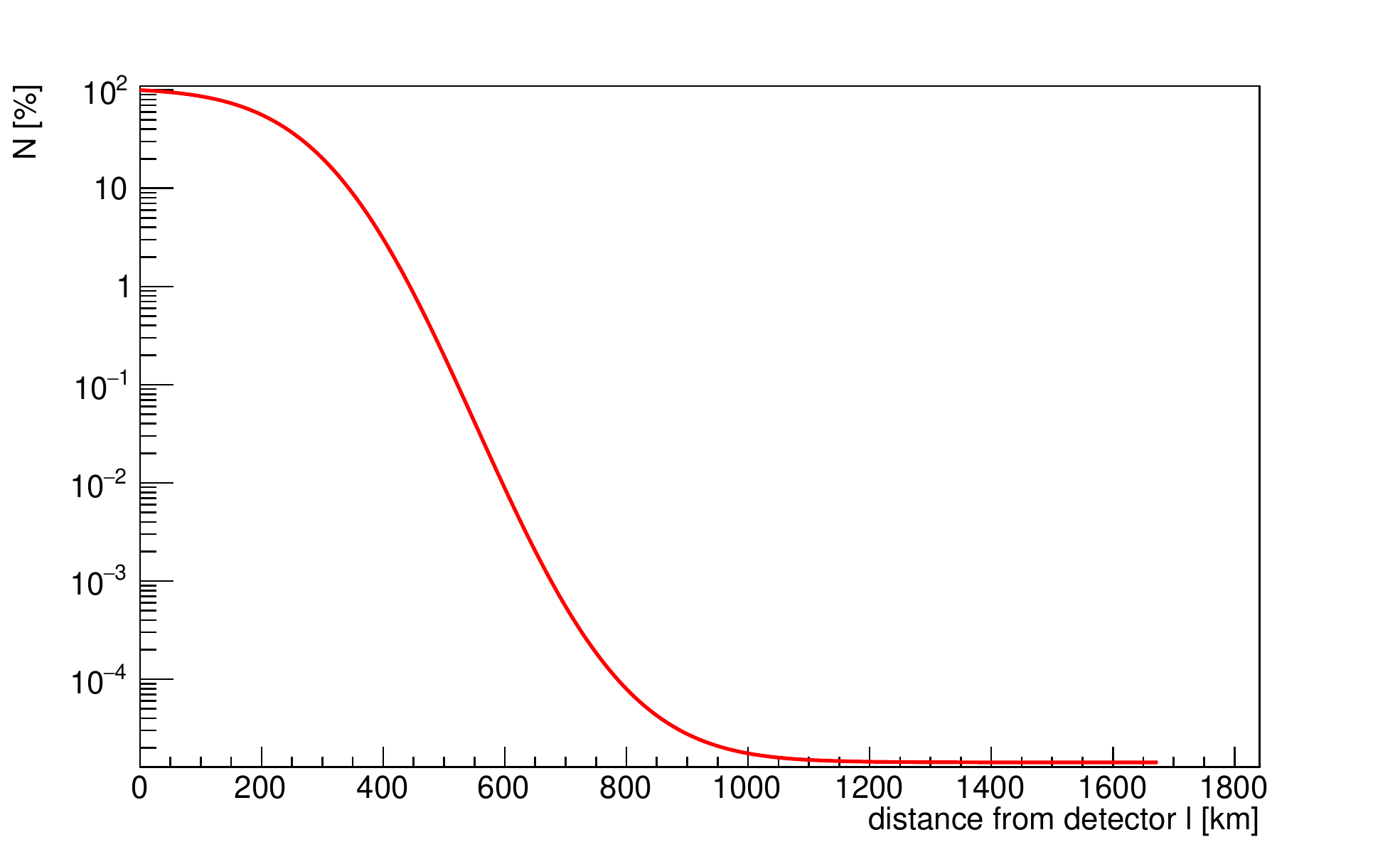}
\caption{Percentage of photons, produced at distance $l$ from the detector, which will reach the detector due to Rayleigh scattering (eq. \ref{eq:Rayleigh}) for detector angle $\psi = 85^{\circ}$ at 38 km altitude.}
\label{fig:Rayleigh}

\end{figure}

Most of UV Cherenkov photons produced in EAS maximum will be scattered on the way to the detector due to Rayleigh scattering. Number of photons $N$ which left from originally emitted $N_0$ after traveling the distance $x$ are:
\begin{equation}
N = N_0 \exp(-\frac{x}{x_0}),
\label{eq:Rayleigh}
\end{equation}
where $x_0 = 0.736\,\mathrm{g/cm^2}$ -- constant determined experimentally connected with mean free path for photon scattering in atmosphere.
Using atmosphere profile from fig. \ref{fig:atmosphere_thickness_profile} and equation \ref{eq:Rayleigh} we can calculate fraction of photons from produced at distance $d$ from SPB2 which will reach the detector (fig. \ref{fig:Rayleigh}). 
Due to observation principle, EUSO detector can not observe light with wavelength shorter than around 300 nm. That is why we do not consider Chrenkov photon absorption by atmospheric ozone.


Analyzing plot from fig. \ref{fig:Rayleigh} we can see that all Cherenkov photons emitted in EAS maximum (about 800 km from detector) will be scattered on the way to detector. Only photons emitted max 400 km from detector can reach it. Only high energetic muons can travel so long distances. For muons decay mean free path  $d_{\mu}$ are described as:
\begin{equation}
d_{\mu} = \gamma\beta c\tau ,
\label{eq:muon_decay}
\end{equation}
where  $\gamma = \frac{E}{m}$ -- Lorentz's factor, $E$ -- muon energy, $m = 105.66\,\mathrm{MeV}$ -- muon mass, $\tau = 2.19\,\mathrm{\mu s}$ -- muon mean lifetime, $\beta = \frac{v}{c}$, $v$ -- muon velocity, $c$ -- light velocity . Considering the minimum decay mean path $d_{min}$ as the distance between first interaction and the point 300 km from SPB2, from where at least 20\% of emitted photons will reach the detector, we can calculate minimum muon energy $E_{min}$ using equation \ref{eq:muon_decay}. For $d_{min} = 800\,\mathrm{km}$ calculated minimum muon energy is $E_{min} \approx 120\,\mathrm{GeV}$. 

\begin{figure}
\centering
\includegraphics[scale=0.5]{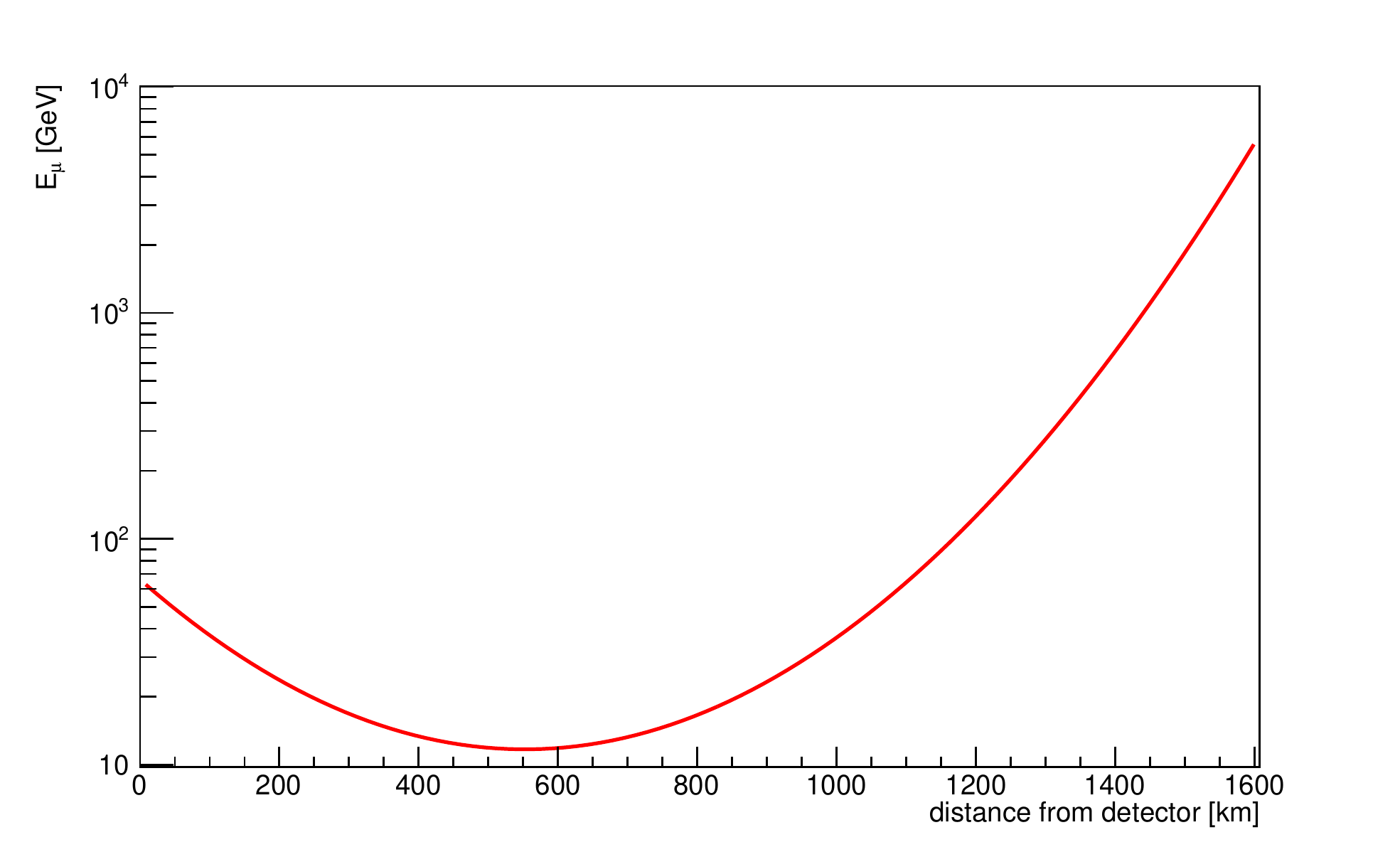}
\caption{Muon energy threshold for UV Cherenkov photons productions in selected location from EUSO-SPB2 detector at 38 km altitude.}
\label{fig:muon_energy_threshold}

\end{figure}

Muon energy threshold for Cherenkov light production in the atmosphere depends on the local atmosphere density. In fig \ref{fig:muon_energy_threshold} muon energy threshold for UV Cherenkov photons production in selected location is plotted. We can see that all the way up to 1200 km from detector the muon energy threshold for Cherenkov production is lower than surviving muon energies (< 120 GeV).

\section{Simulations results}
Taking into consideration all things described above we can calculate Cherenkov photons distribution close to SPB2 detector using reasonable approximation. EAS will fully develop 800 km from the detector and almost all UV photons emitted at shower maximum will be scattered.
Only muons with minimum energy $E_{min} = 120\,\mathrm{GeV}$ can travel close to detector. Running vertical EAS simulations we can get high energy muon distribution at ground level, i.\,e. muon distribution after developing in $1000\,\mathrm{g/cm^2}$ of atmosphere. We can use this distribution as distribution for horizontal EAS after traveling the same amount of the atmosphere. Having from simulations all information about muons' positions and momenta we can calculate muon distributions 300 km from the detector (the point from which 20\% of emitted photons will reach the detector). Taking into account muon distribution, atmosphere profile on the way to detector and Rayleigh scattering we can calculate Cherenkov photons distributions around the detector. 

\begin{figure}
\includegraphics[scale=0.4]{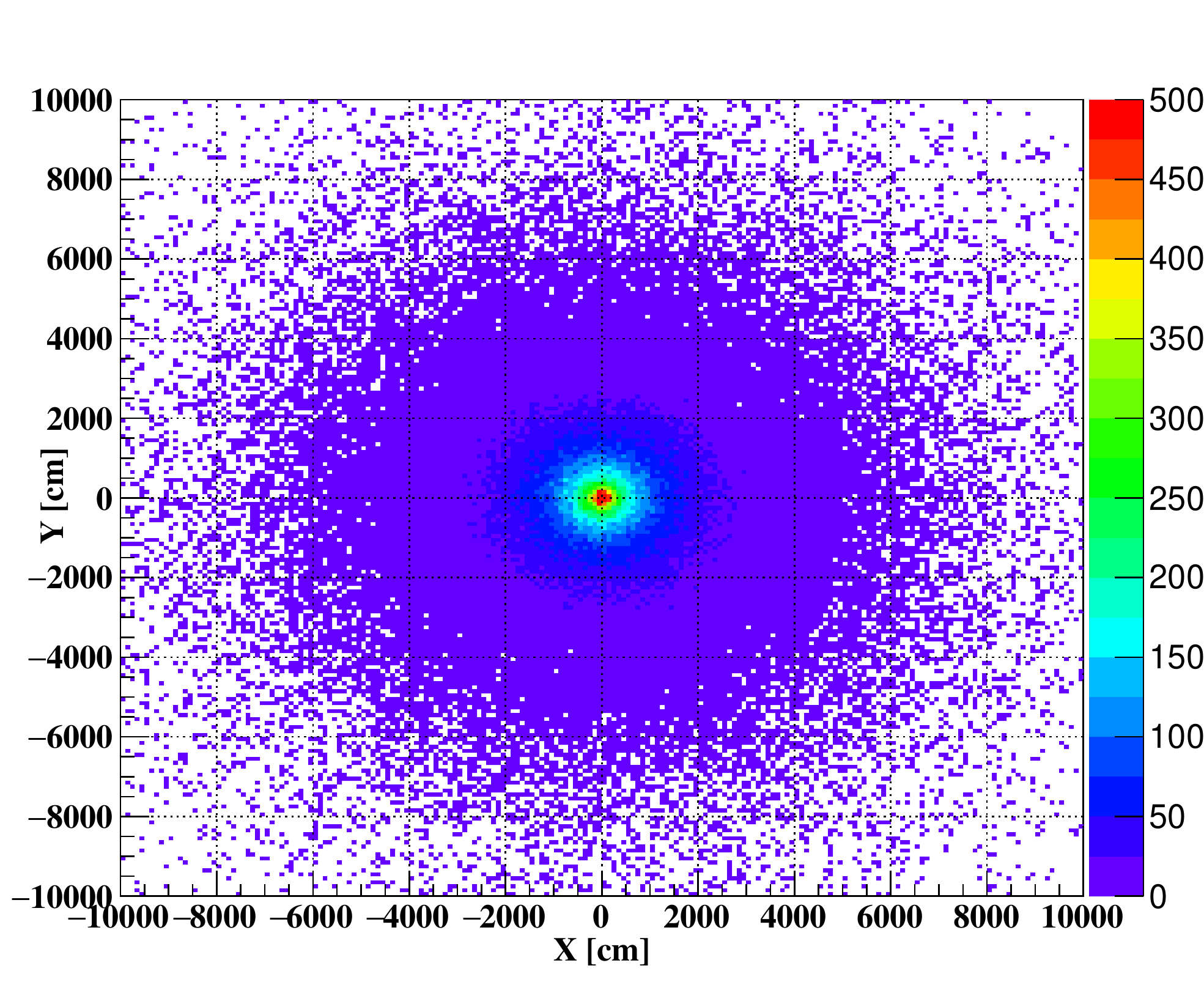}
\includegraphics[scale=0.4]{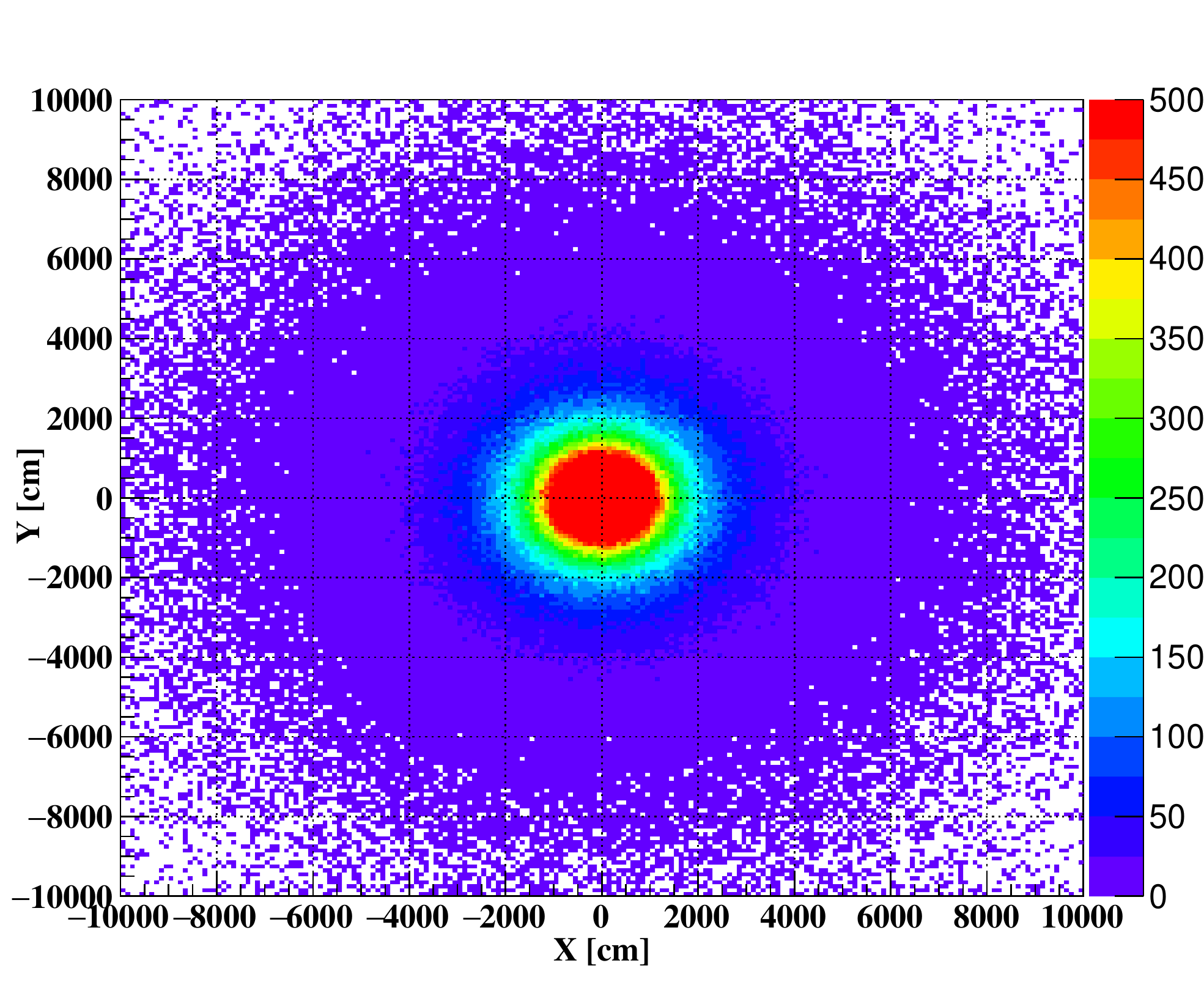}
\caption{Muon distribution on ground level obtained from simulations of 100 vertical EAS induced by $10^{17}\,\mathrm{eV}$ (left) and $10^{18}\,\mathrm{eV}$ (right) proton. }
\label{fig:muon_distribution}

\end{figure}

Muon lateral distributions on ground level obtained from simulations of 100 vertically developing EAS induced by $10^{17}\,\mathrm{eV}$ and $10^{18}\,\mathrm{eV}$ protons are plotted in fig. \ref{fig:muon_distribution}. Number of produced muons increase with primary particle energy and muon distribution's width changes only slightly. Muon lateral distribution angle depends on ratio of average transverse momenta (about $0.5\,\mathrm{GeV/c}$) and muon momentum (min. 120 GeV), so their trajectories are not far from EAS axis.

\begin{figure}
\centering
\includegraphics[scale=0.5]{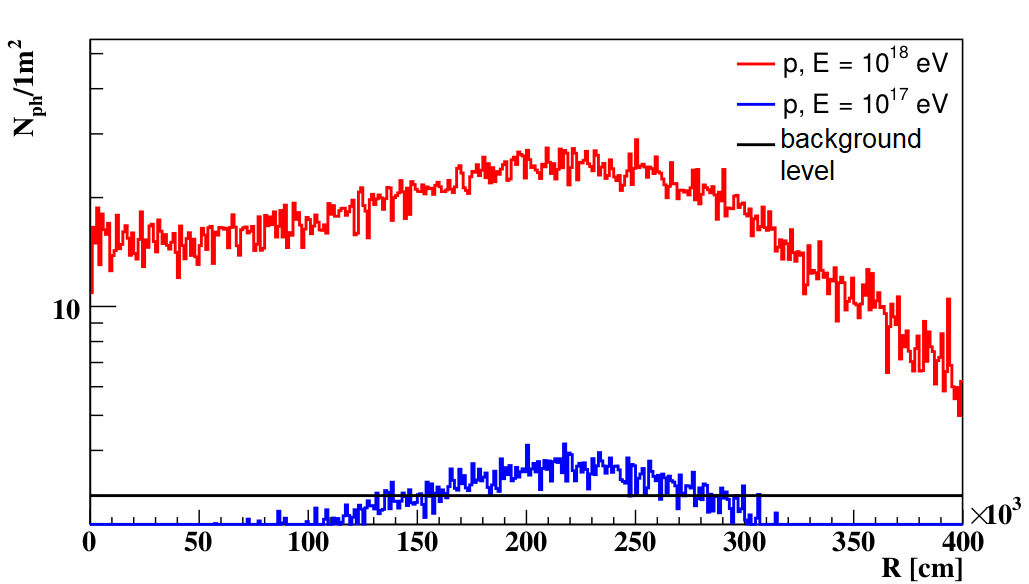}
\caption{Number of Cherenkov photons $N_{ph}$ per $1\,\mathrm{m^2}$ produced by muons of a single average EAS, in function of distance $R$ from shower axis, in EAS induced by $10^{17}\,\mathrm{eV}$ (blue line) and  $10^{18}\,\mathrm{eV}$ (red line) proton.}
\label{fig:photon_distribution}

\end{figure}

Calculated from muons distributions number of Cherenkov photons produced on $1\,\mathrm{m^2}$ around detector are shown in fig. \ref{fig:photon_distribution}.

EUSO-SPB2 detector will be detecting photons with $~1\,\mathrm{m^2}$  effective area. In the fig. \ref{fig:photon_distribution} we can see that up to $3200\,\mathrm{m}$ from EAS axis number of photons per $1\,\mathrm{m^2}$ for EAS generated by $10^{17}\,\mathrm{eV}$ proton are almost constant. It gives $A = 3.2\cdot 10^7 \,\mathrm{m^2}$ area in with EUSO-SPB2 detector must be to detect UV Cherenkov photons. For $10^{18}\,\mathrm{eV}$ proton number of photons per $1\,\mathrm{m^2}$ are higher and the area of observation is bigger: $A \approx 5\cdot 10^7 \,\mathrm{m^2}$ (for 5 photons or more at $1\,\mathrm{m^2}$).

SPB2 Cherenkov detector will be looking on $1^{\circ}$ at horizon with $40^{\circ}$ width. It gives $\Omega = 0.012\,\mathrm{sr}$ observation solid angle. SPB balloons are nominally fly for about 100 days, but detector can works only at night. It gives about  $t_{obs} = 2.6 \cdot 10^6\,\mathrm{s}$ (30 days) of observation time. Calculating geometric factor $GF$:
\begin{equation}
GF = A\cdot \Omega \cdot t_{obs}
\end{equation}
and $GF$ are $1.0\cdot 10^{12}\,\mathrm{m^2 sr\, s}$ and 
$1.6\cdot 10^{12}\,\mathrm{m^2 sr\, s}$ respectively for $10^{17}\,\mathrm{eV}$
 and $10^{18}\,\mathrm{eV}$ proton. Integrating cosmic rays flux  \cite{blumer2009cosmic} from $10^{17}\,\mathrm{eV}$ and $10^{18}\,\mathrm{eV}$ up to the highest energies we get exposure $1.2\cdot 10^{-10} \mathrm{m^{-2} sr^{-1} s^{-1}}$ and $1.2\cdot 10^{-12} \mathrm{m^{-2} sr^{-1} s^{-1}}$ respectively. 
From this we can expect to receive 120 events (min. 3 UV Cherenkov photons$/\mathrm{m^2}$) for CR with energies above $10^{17}\,\mathrm{eV}$  and about 1 event (min. 5 UV Cherenkov photons$/\mathrm{m^2}$) for energies above $10^{18}\,\mathrm{eV}$.

\section{Conclusion}
The aim of this work was to check possibility of detecting horizontal Extensive Air Showers by UV Cherenkov photons in range 290 nm - 400 nm hitting directly into EUSO-SPB2 detector. To obtain Cherenkov signal we had to run Monte Carlo Simulations of EAS development. For EAS developing almost horizontally, distance between the beginning of the atmosphere and SPB2 detector is about 1600 km long, and thickness of the atmosphere is about 10 times bigger than for vertically developing showers. Most of particles produced in shower maximum will be scattered and only muons with energy higher than $120\,\mathrm{GeV}$ can reach the detector. Due to Rayleigh  scattering UV Cherenkov photons produced in shower maximum will be scattered and only signal from photons emitted by very energetic muons gives the chance of EAS detection. 

Running vertical EAS simulations we get the distribution of muons with energy $120\,\mathrm{GeV}$ or higher produced in shower developing in 1 kg of atmosphere. Number of high energetic muons in horizontal EAS should be higher because such shower develops in lower densities, however a significant fraction of them would decay in a long way to the detector. Work on more accurate simulations are in progress. 

Having muon distribution 800 km from SPB2 detector we could calculate how many UV Cherenkov photons will be produced as muons will be traveling to the balloon-borne instrument. Knowing UV Cherenkov photons' distribution, cosmic ray flux, parameters of detector and detection time we can estimate number of expected events. For 100 days of flight we expect to observe 120 events for CR with energies above $10^{17}\,\mathrm{eV}$  and about 1 event for energies above $10^{18}\,\mathrm{eV}$.

\section*{Acknowledgments}
This work was supported by grant 2017/27/B/ST9/02162 funded by National Science Centre in Poland.

\bibliography{bibliografia}

\end{document}